\documentclass[12pt]{article}

\usepackage{geometry}
\usepackage{amssymb,amsmath,amsthm}
\usepackage{graphicx}
\usepackage{ dsfont }
\newcommand{\Var} {\mbox{$\rm{Var}$\,}}
\newcommand{\Prob} {\mbox{$\rm{Prob}$\,}}

\begin{document}

\title{An {\tt R} Implementation of the P\'olya-Aeppli Distribution}

\author{Conrad J. Burden\\
Mathematical Sciences Institute\\
Australian National University\\
Canberra, A.C.T. 0200, Australia}

\maketitle

\begin{abstract}
An efficient implementation of the P\'olya-Aeppli, or geometirc compound Poisson, distribution in the statistical programming language {\tt R} 
is presented.  The implementation is available as the package {\tt polyaAeppli} and 
consists of functions for the mass function, cumulative distribution function, quantile function and random variate generation 
with those parameters conventionally provided for standard univatiate probability distributions in the  {\tt stats} package in {\tt R}
\end{abstract}



\section[Introduction]{Introduction}

A P\'olya-Aeppli or geometric compound Poisson random variable is defined as the sum of a Poisson number of 
independent and identically distributed geometric random variables~\cite{Polya30,Johnson92}.  
Its distribution arises as an approximation to the distribution of the number of occurrences of a given short word in a random Markovian sequence of letters from a finite 
alphabet~\cite{Schbath95,Robin01}, and as an approximation to the number of word matches between two random sequences~\cite{Lippert02}.  
Thus the Polya-Aeppli distribution finds applications in bioinformatics in searches for over- or underrepresented words, which may signal functional elements within genomes, 
and applications associated with alignment-free measures of sequence similarity.   
It is also appropriate as an alternative to the commonly used negative binomial model of over-dispersed Poisson count data which arises, for instance, from high throughput 
sequencing experiments~\cite{Esnaola13}.  

The {\tt stats} package in {\tt R}~\cite{R01} contains implementations of many standard univariate probability distributions as functions for the density/mass function, 
cumulative distribution function, quantile function and random variate generation.  We present here an efficient implementation of these 4 functions for the P\'olya-Aeppli distribution, 
available as the package {\tt polyaAeppli}~\cite{Burden13b}. 
The implementation relies on iterative formulae for the log of the mass and cumulative distribution functions, which allow an accurate evaluation of the distribution in the extreme 
upper and lower tails.  

We define a P\'olya-Aeppli random variable as 
\begin{equation}
X = \sum_{i = 1}^N Y_i	\label{PAdefinition}
\end{equation}
where $N$ is a Poisson random variable with parameter $\lambda$ with probability mass function 
\begin{equation}
P_N(n; \lambda) = \Prob(N = n) = \frac{e^{-\lambda} \lambda^n}{n!}, \qquad n = 0, 1, \ldots, 
\end{equation}
and the $Y_i$ are identically and independently distributed shifted geometric random variables with parameter $p$ with common probability mass function 
\begin{equation}
P_Y(y; p) = \Prob(Y = y) = p^{y - 1} (1 - p), \qquad y = 1, 2, \ldots.   
\end{equation}
The P\'olya-Aeppli probability mass function is~\cite{Johnson92} 
\begin{eqnarray}
P_X(x; \lambda, p) & = & \Prob(X = x)  \nonumber \\  \nonumber \\
	& = & \begin{cases} 
		e^{-\lambda} & \mbox{if $x = 0$} \\
		e^{-\lambda} \displaystyle\sum_{n = 1}^y \frac{ \lambda^n}{ n!} {{y - 1}\choose{n - 1}} p^{y - n} (1 - p)^n &  \mbox{if $x = 1, 2, \ldots$.}
		\end{cases}			\label{massFunc}
\end{eqnarray}
The mean and variance are 
\begin{equation}
E(X) = \mu = \frac{\lambda}{1 - p}, \quad \Var(X) = \sigma^2 = \lambda \frac{1 + p}{(1 - p)^2}.  
\end{equation}
These formulae invert to give 
\begin{equation}
\lambda = \frac{2 \mu^2}{\sigma^2 + \mu}, \quad p = \frac{\sigma^2 - \mu}{\sigma^2 + \mu}. 
\end{equation}
Note that the restriction $0 \le p <1$ implies that $\sigma^2 > \mu$, making ths distribution a suitable model for over-dispersed count data.  
When $p = 0$, the P\'oya-Aeppli distribution reduces to the Poisson distribution with mean $\lambda$.

\section[Implementation]{Implementation}

\begin{figure}[t!]
\begin{center}
		\includegraphics[width=0.55\textwidth]{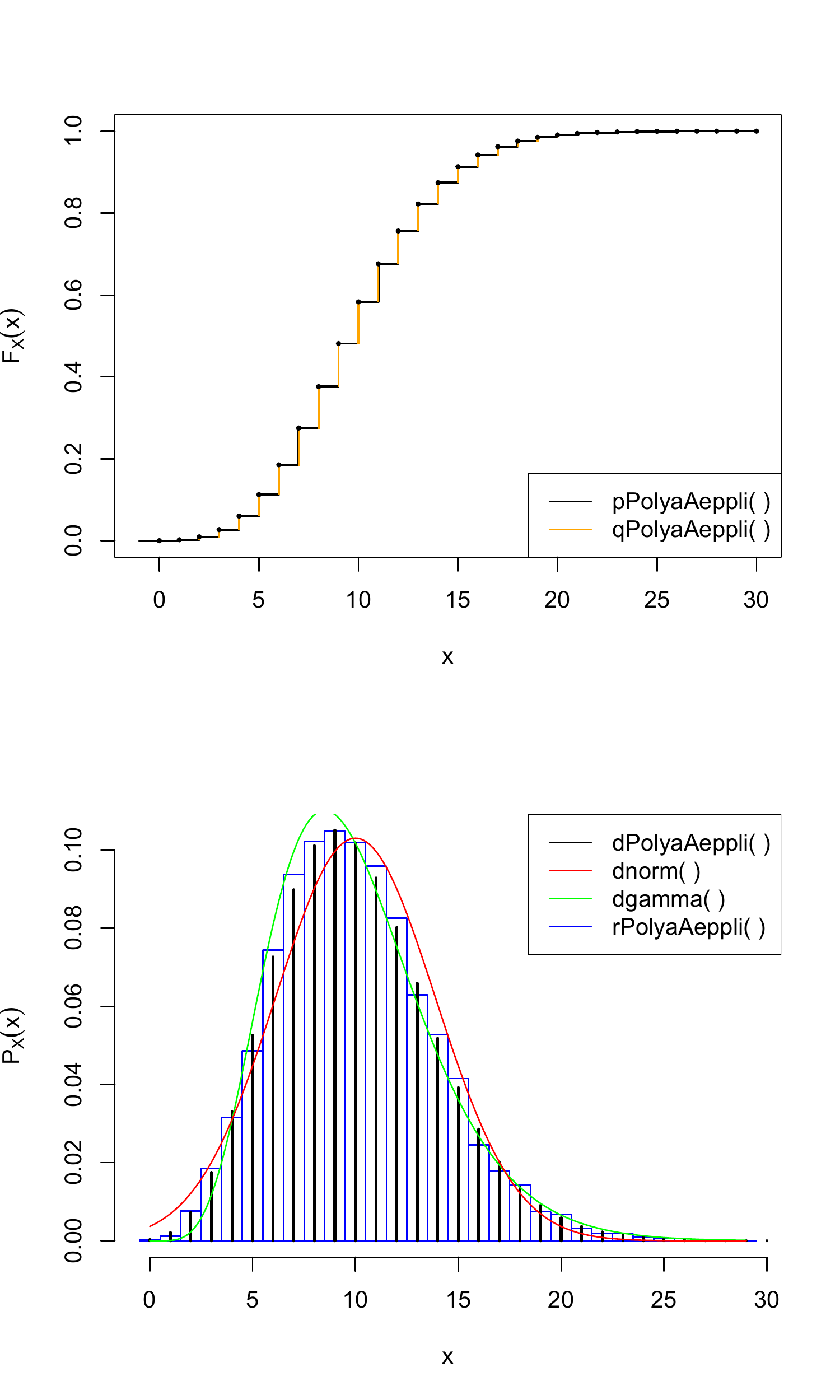}
\caption{The P\'olya-Aeppli distribution for values of $\lambda$ and $p$ corresponding to mean $\mu = 10$ and variance $\sigma^2 =15$.  
Top panel: the cumulative distribution calculated from {\tt pPolyaAeppli()} (black) and quantiles calculated from 
{\tt qPolyaAeppli()} plotted on the horizontal axis against probabilites in the range $[0, 1)$ plotted on the vertical axis (orange).  
Bottom panel: The probability mass function calculated from {\tt dPolyaAeppli()} (black) and a histogram of 10\,000 random variables 
sampled using {\tt rPolyaAeppli()}.  Also plotted for comparison are normal (red) and gamma (green) distributions with matching mean and variance.  }  
\label{fig:ExampleDistrib}
\end{center}
\end{figure}

Consistent with the conventions used in {\tt R}, our implementation comprises the four functions \\ \\
{\tt dPolyaAeppli(x, lambda, prob, log = FALSE)} \\
{\tt pPolyaAeppli(q, lambda, prob, lower.tail = TRUE, log.p = FALSE)}\\
{\tt qPolyaAeppli(p, lambda, prob, lower.tail = TRUE, log.p = FALSE)} \\
{\tt rPolyaAeppli(n, lambda, prob)} \\ \\
for the mass function, cumulative distribution function, quantile function and random variate generation respectively.  The arguments 
{\tt lambda} and {\tt prob} accept single values for the parameters $\lambda$ and $p$ respectively.  The argument {\tt x} is a 
vector of (non-negative  integer) quantiles, {\tt q} is a vector of quantiles, {\tt p} is a vector of probabilities and {\tt n} is the number of 
random values to return.  If the logical variables {\tt log} or {\tt log.p} are {\tt TRUE}, probabilities are given as logarithms to base $e$.  
If the logical variable {\tt lower.tail} is set to its default value of {\tt TRUE} probabilities are $\Prob(X \le x)$, otherwise probabilities are 
$\Prob(X > x)$.  Examples of plots of the cumulative distribution, mass function and a histogram of a random sample are shown in Figure~\ref{fig:ExampleDistrib} 
for values of $\lambda$ and $p$ corresponding to a mean $\mu = 10$ and variance $\sigma^2 =15$.  

\begin{figure}[t!]
\begin{center}
		\includegraphics[width=0.85\textwidth]{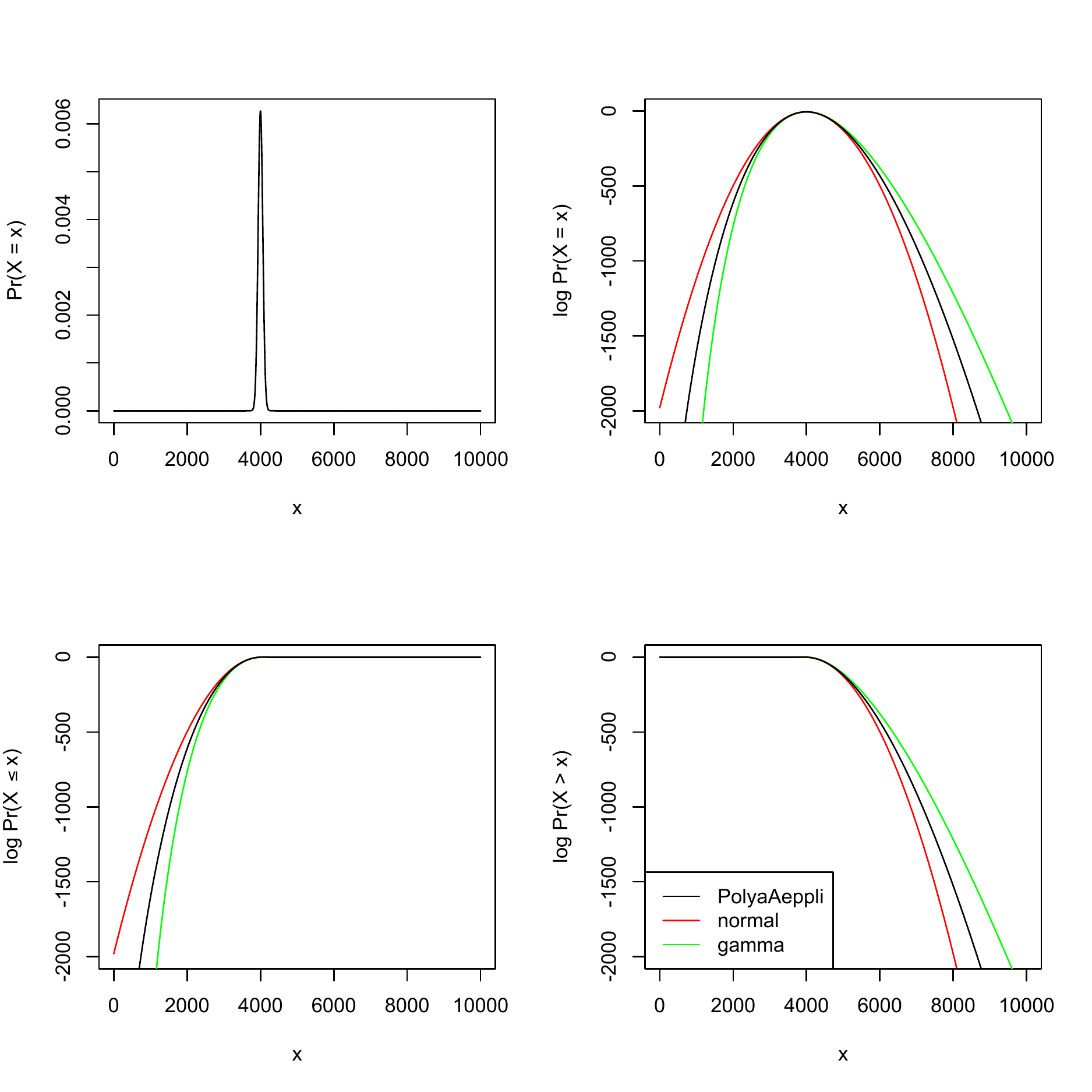}
\caption{Extreme tails of the P\'olya-Aeppli distribution for values of $\lambda$ and $p$ corresponding to mean $\mu = 4000$ and variance $\sigma^2 =4050$.  
Top left: The mass function calculated using {\tt dPolyaAeppli()} with defaults.  Top right: Log of the mass function using {\tt dPolyaAeppli(\ldots, log=TRUE)}.  
Bottom left: Log of the lower tail of the cumulative distribution calculated using {\tt pPolyaAeppli(\ldots, log.p=TRUE)}.  
Bottom right: Log of the upper tail of the cumulative distribution calculated using {\tt pPolyaAeppli(\ldots, log.p=TRUE, lower.tail = FALSE)}.  
Also plotted for comparison are normal (red) and gamma (green) distributions with equivalent parameters. }  
\label{fig:Extreme_tail_plots}
\end{center}
\end{figure}

The option {\tt log = TRUE} allows for a more precise determination of the distribution in the extreme tails when {\tt dPolyaAeppli()} would 
otherwise return 0 to machine accuracy.  The option {\tt log.p = TRUE} allows for an accurate determination of the cumulative distribution 
using {\tt pPolyaAeppli()} and its inverse using {\tt qPolyaAeppli()} in the extreme lower and upper tails when used in conjunction with {\tt lower.tail = TRUE}
and  {\tt FALSE} respectively.  Plots of the mass function and cumulative distribution over a range of quantiles more than 60 standard deviations 
either side of the mean for values of $\lambda$ and $p$ corresponding to mean $\mu = 4000$ and variance $\sigma^2 =4050$ are shown in 
Figure~\ref{fig:Extreme_tail_plots}.   

{\tt R} implementations of the P\'olya-Aeppli mass function and random variate generator, but not the cumulative distribution function or quantile function, 
currently exist via the Poisson-Tweedie distribution in the package {\tt tweeDEseq}~\cite{Esnaola13}.  
The function {\tt dPT(x, mu = lambda/(1 - prob), D = (1 + p)/(1 - p), a = -1)} plays the role of {\tt dPolyaAeppli(x, lambda, prob)}, and the 
function\\ {\tt rPT(n, mu = lambda/(1 - prob), D = (1 + p)/(1 - p), a = -1)} plays the role of {\tt rPolyaAeppli(n, lambda, prob)}.  However, because the {\tt log = TRUE} 
option is not implemented, {\tt dPT()} is not able to evaluate the distribution as accurately as {\tt dPolyaAeppli(\ldots, log = TRUE)} in the extreme tails of the distribution.  
Furthermore the {\tt tweeDEseq} implementations are not vectorised  over the parameters of the distribution.  

An {\tt R} implementation of the upper tail of the P\'olya-Aeppli distribution was previously implemented via the now discontiued function {\tt computePValue()} 
in the package {\tt NeMo}~\cite{Picard08} for the study of network motifs.  This function, with arguments   
{\tt computePValue(lambda, prob, 1+x)} played the same role as 
{\tt pPloyaAeppli(x, lambda, prob, lower.tail=FALSE)}, but was not vectorised over either the argument {\tt x} or the parameters {\tt lambda} or {\tt prob}.  
Furthermore the option {\tt log.p=TRUE}, was not implemented, making the extreme tails of the distribution inaccessible.  

\section[Algorithms]{Algorithms}

The implementation of the P\'olya-Aeppli distribution must be able to evaluate the mass function and cumulative distribution function efficiently for a range of input quantiles.  
From Eq.~(\ref{massFunc}) Evens~\cite{Evans53} has derived the recurrence formula 
\begin{eqnarray}
\lefteqn{(x + 1) P_X(x + 1; n, \lambda) = } \nonumber \\
& & [\lambda(1 - p) + 2px] P_X(x; n, \lambda) - p^2 (x - 1) P_X(x - 1; n, \lambda), \quad x = 1, 2, \ldots,     \label{recurrence}
\end{eqnarray}
with 
\begin{equation}
P_X(0; n, \lambda) =  e^{-\lambda}, \quad P_X(1; n, \lambda) = \lambda (1 - p) e^{-\lambda}. 
\end{equation}
As is stands this formula is not adequate for our purposes as it susceptible to roundoff errors when $e^{-\lambda}$ is small, which may happen for instance when $\mu$ is 
moderately large and $\sigma^2/\mu$ close to 1.  Nuel~\cite{Nuel08} notes that 
this problem is overcome by calculating instead the logarithm of the mass function 
\begin{equation}
l(x) = \log P_X(x; n, \lambda), 
\end{equation}
from the recurrence relation
\begin{gather}   \label{logRecurrence}
l(x + 1) = \log\left\{ \frac{\lambda (1 - p) + 2px - p^2(x - 1) e^{l(x - 1) - l(x)}}{x + 1} \right\} + l(x), \quad x = 1, 2, \ldots, \\
l(0) = -\lambda, \quad l(1) = -\lambda  + log \lambda(1 - p),  
\end{gather}
which is easily derived from Eq.(\ref{recurrence}).   
The function {\tt dPolyaAeppli(x, lambda, prob)} is calculated by first evaluating this iterative formula out to the largest finite value in the array {\tt x} and 
exponentiating if {\tt log=FALSE}.  

To calculate the lower tail of cumulative distribution 
\begin{equation}
F_X(x) = \Prob(X \le x) = \sum_{i = 0}^{\lfloor x \rfloor} P_X(i), \quad x \in \mathds{R}, 
\end{equation}
Nuel~\cite{Nuel08} similarly avoids serious roundoff errors by calculating the logarithm of the cumulative distribution function
\begin{equation}
g(x) = \log F_X(x), 
\end{equation}
from the recurrence relation
\begin{gather} \label{logCumulRecurrence}
g(i + 1) = g(i) + \log\left(1 + e^{l(i + 1) - g(i)} \right), \quad i = 0, 1, 2, \ldots \\
g(0) = \lambda.  
\end{gather}
The function {\tt pPolyaAeppli(x, lambda, prob, lower.tail = TRUE)} is calculated by first evaluating Eqs.~(\ref{logRecurrence}) and (\ref{logCumulRecurrence}) 
out to the integer part of the largest finite value in the array {\tt x} and exponentiating if necessary.  
 
This procedure will not meet the required accuracy for the extreme upper tail.  In this case we consider the log of the upper tail, 
\begin{equation}
h(x) = \log(1 - F_X(x)) = \log \Prob(X > x), 
\end{equation}
which can be iterated downwards via the formula 
\begin{equation} \label{logDownwardRecurrence}
h(i - 1) = h(i) + \log(1 + e^{l(i) - h(i)}), \quad i = \lfloor x_{\rm max}\rfloor, \lfloor x_{\rm max}\rfloor - 1, \lfloor x_{\rm max} \rfloor - 2, \ldots, 1.  
\end{equation}
To evaluate {\tt pPolyaAeppli(x, lambda, prob, lower.tail = FALSE)} where {\tt x} includes finite values greater than the mean $\mu$, 
$x_{\rm max}$ is taken to be the largest finite value in the array {\tt x}, and Eq.(\ref{logDownwardRecurrence}) 
is initiated by calculating 
\begin{equation}
h(\lfloor x_{\rm max}\rfloor) = l(\lfloor x_{\rm max}\rfloor + 1) + \log \sum_{i = \lfloor x_{\rm max}\rfloor  + 1}^\infty e^{l(i) - l(\lfloor x_{\rm max}\rfloor  + 1)}, 
\end{equation}
which converges to machine accuracy reasonably rapidly.  

The functions $l(x)$, $g(x)$ and $h(x)$ are calculated by auxilliary functions\\ {\tt lPolyaAeppliArray()},  {\tt gArray()} and {\tt hArray()}, and $h(x_{\rm max})$ 
is evaluated by the auxilliary function {\tt logTail()}.  

The quantile function {\tt qPolyaAeppli()} is evaluated by first evaluating probabilities with {\tt pPolyaAeppli()} from {\tt q} $=0$ out to a safe upper bound, 
and seeking the maximum {\tt q} entailing a probability less than the specified probability for each value in the input array {\tt p}.  The upper bound used is the value 
of he quantile function {\tt qgamma()} of the gamma distribution with matching mean and variance (see Fig.~\ref{fig:Extreme_tail_plots}) evaluated at the maximum finite value in the 
input array {\tt p}, plus one standard deviation.  

The function {\tt rPolyaAeppli()} is evaluated directly from the definition Eq.(\ref{PAdefinition}) via the {\tt R} functions {\tt rpois()} and {\tt rgeom()}.  

\section*{Acknowledgments}

CJB acknowledges valuable assitance from John Maindonald related to generating a package from the {\tt R} code described herein, and support from NHMRC Grant 525453. 

\bibliographystyle{plain}

\end{document}